\documentclass[aps,prl,twocolumn,superscriptaddress,showpacs]{revtex4}

\usepackage{graphicx}

\begin{document}


\title{Statics and dynamics of an incommensurate spin order in a geometrically frustrated antiferromagnet CdCr$_2$O$_4$}


\author{J.-H. Chung}
\affiliation{NIST Center for Neutron Research, National Institute
of Standards and Technology, Gaithersburg, Maryland 20899-8552,
USA}
\author{M. Matsuda}
\affiliation{Advanced Science Research Center, Japan Atomic Energy
Research Institute, Tokai, Ibaraki 319-1195, Japan}
\author{S.-H. Lee}
\affiliation{Department of Physics, University of Virginia,
Charlottesville, Virginia 22904, USA}
\author{K. Kakurai}
\affiliation{Advanced Science Research Center, Japan Atomic Energy
Research Institute, Tokai, Ibaraki 319-1195, Japan}
\author{H. Ueda}
\author{T.J. Sato}
\affiliation{Institute for Solid State Physics, University of
Tokyo, Kashiwa, Chiba 277-8581, Japan}
\author{H. Takagi}
\affiliation{Graduate School of Frontier Science, University of
Tokyo, Kashiwa, Chiba 277-8581, Japan}
\author{K.-P. Hong}
\author{S. Park}
\affiliation{HANARO Center, Korea Atomic Energy Research
Institute, Daejeon, Korea}

\date{\today}

\begin{abstract}
Using elastic and inelastic neutron scattering we show that a
cubic spinel, CdCr$_2$O$_4$, undergoes an elongation along the $c$-axis
($c > a = b$) at its spin-Peierls-like phase transition at $T_N$ =
7.8 K. The N\'{e}el phase ($T < T_N$) has an incommesurate spin
structure with a characteristic wave vector \textbf{Q}$_M$ =
(0,$\delta$,1) with $\delta \sim$ 0.09 and with spins lying on the 
$ac$-plane. 
This is in stark contrast to another well-known Cr-based
spinel, ZnCr$_2$O$_4$, that undergoes a $c$-axis contraction and a
commensurate spin order. 
The magnetic excitations of the
incommensurate N\'{e}el state has a weak anisotropy gap of 0.6 meV and
it consists of at least three bands extending up to 5 meV. 
\end{abstract}


\maketitle



When a transition metal oxide has degeneracy, orbital or magnetic,
a novel phase transition can occur through the coupling of the
relevant degrees of freedom to the lattice to lift the degeneracy.
A well-known example is the Jahn-Teller lattice distortion
that involves either doubly degenerate $e_g$ orbitals or triply
degenerate $t_{2g}$ orbitals \cite{jt,tokura}. Geometrically frustrated magnets
provide a fertile ground for similar novel phase transitions
because the degeneracy can be macroscopic both for quantum and for classical spins \cite{frust}. Until now the
most frustrating system known is the one that consists of a network of corner-sharing
tetrahedra, the pyrochlore lattice, with the
simplest spin hamiltonian, ${\cal H}_1=J\sum_{NN}S_i\cdot S_j$
with isotropic antiferromagnetic nearest neighbor interactions
only. Theoretically, it has been shown that in the ideal case, the spins
alone cannot order even at zero temperature
\cite{moes98,cana00}. Experimentally the magnetic lattice can be
realized in several materials such as in the
pyrochlores A$_2$B$_2$O$_7$ \cite{ging}, spinels
AB$_2$O$_4$ \cite{lee00} and C15 Lave phases AB$_2$ \cite{ball96}. Among them, 
Cr-based spinels ACr$_2$O$_4$ (A=Zn, Cd) realize the most frustrating 
lattice with the dominant antiferromagnetic nearest neighbor interactions due to the direct overlap of the 
$t_{2g}$ orbitals of the neighboring Cr$^{3+} (3d^3)$ ions \cite{good60,samu70}. 
Consequently, ACr$_2$O$_4$ remains paramagnetic to
temperatures far below the characteristic strength of the interactions
between the spins, the Curie-Weiss temperature $|\Theta_{CW}|=$
390 K and 88 K for A = Zn \cite{frust} and Cd \cite{meny66,rove02,ueda05}, respectively.
Upon further cooling, however, the system undergoes a first order
spin-Peierls-like phase transition \cite{lee00,sjt} from a cubic paramagnet to 
a tetragonal N\'{e}el state at $T_N=$ 12.5 K and 7.8 K for A = Zn \cite{lee00}
and Cd \cite{rove02, ueda05}, respectively. 

Recently, ZnCr$_2$O$_4$
has been studied extensively using neutron scattering techniques \cite{lee00,lee02}. Its tetragonal distortion involves a {\it
contraction} along the $c$ axis ($c < a$). Its N\'{e}el state has a
rather complex {\it commensurate} spin structure. The
spin structure has four different characteristic wave vectors, 
\textbf{Q}$_M$, 
($\frac{1}{2}$,$\frac{1}{2}$,0), (1,0,$\frac{1}{2}$),
($\frac{1}{2}$,$\frac{1}{2}$,$\frac{1}{2}$) and (0,0,1) \cite{leeun}.
Furthermore, the relative ratios of the neutron scattering
intensities of these wave vectors vary depending on the subtle chemical
conditions during sample preparation \cite{leeun}. 
This suggests that even in the tetragonal phase,
ZnCr$_2$O$_4$ is critically located close to several spin
structures, that makes it difficult to understand the true
nature of its ground state.
The CdCr$_2$O$_4$ compound in this class has not received much attention partly because it was commonly believed that the same physics hold true as 
in ZnCr$_2$O$_4$ \cite{rove02}, which we found is not the case.
As the pyrochlore lattice possesses many ground states and multiple ways for 
lifting the ground state degeneracy, CdCr$_2$O$_4$ provides a venue for 
understanding phase transitions in this seemingly complex class of materials.

In this paper, we report the results from elastic and inelastic neutron scattering measurements on $^{114}$CdCr$_2$O$_4$ (space group $Fd\bar{3}m$, $a=8.58882$ \AA~ for $T=$ 10 K). Surprisingly, we find that it undergoes a phase transition that is qualitatively
different in nature from the one observed in ZnCr$_2$O$_4$. CdCr$_2$O$_4$
{\it elongates} along the $c$-axis ($c>a$) and undergoes an {\it
incommensurate} (IC) N\'{e}el order.
The high \textbf{Q}-resolution data indicate that the
incommensurate magnetic structure has a {\it single} characteristic wave
vector of \textbf{Q}$_M=(0,\delta,1)$ with 
$\delta\sim$ 0.09 perpendicular to the unique $c$-axis. 
We present two possible high symmetry spin structures that are consistent with the particular \textbf{Q}$_M$. The interplay between the lattice
distortion and the IC spin structure is discussed
along with dispersion of the spin wave excitations. 
%

Preliminary measurements were initially performed on the TAS2 thermal 
triple-axis spectrometer of Japan Atomic Energy Research Institute (JAERI), 
while more detailed studies ensued on 
the SPINS cold neutron triple-axis spectrometer of NIST Center for Neutron Research.  A single crystal weighing $\sim$ 100 mg was used for the elastic
measurements, while three of
these crystals were co-mounted within 1$^o$ mosaic for the inelastic
measurements. The
crystals were mounted in the ($hk$0) scattering plane, that
allowed the investigations of three equivalent planes, ($hk$0),
(0$kl$), and ($h$0$l$) planes due to the crystallographic domains.

\begin{figure}
 \includegraphics[width=2.8in, angle=90]{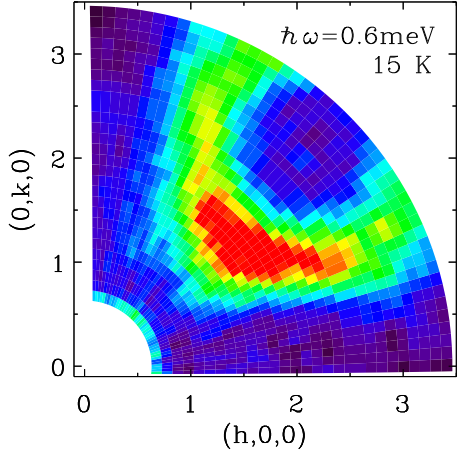}
 \caption{\label{ring}Color image of inelastic neutron scattering intensities from single crystals of CdCr$_2$O$_4$ at $T$ = 15 K $> T_N$ with $\hbar\omega$ = 0.6 meV. The data were taken at SPINS using eleven 2.1 cm x 15 cm PG(002) analyzer blades with $E_f=$ 5meV.}
\end{figure}
\begin{figure}
 \includegraphics[width=3.6in]{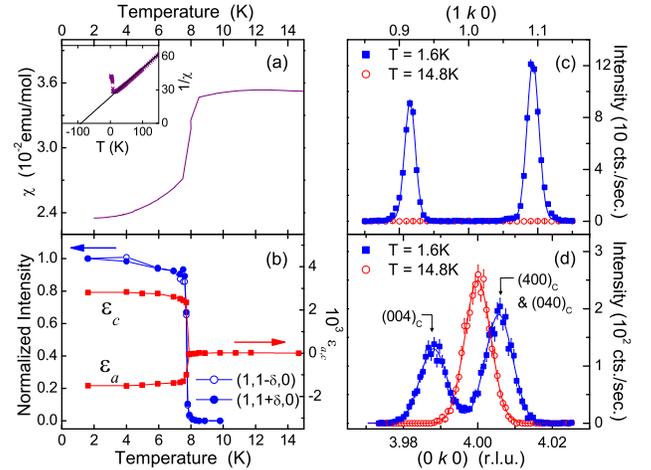}
 \caption{\label{data1} (a) Bulk susceptibility, $\chi$,
 as a function of $T$. The transition is of first-order and has a small but finite hysteresis over 0.1 degrees. The inset shows a linear
 fit to $1/\chi$. (b) $T$-dependence of the
 normalized integrated neutron scattering intensity at the magnetic peaks and that of 
the lattice strains, $\epsilon_{a,c}=\frac{\Delta (a,c)}{(a,c)}$.
The data were obtained from fitting the data in (c) and (d) to gaussians. (c), (d) Elastic neutron scattering data (c) through magnetic (1,$\delta$,0) IC points and (d) through a nuclear (400) Bragg reflection below and above $T_N$.}
\end{figure}

Fig. \ref{ring} shows the $\bf Q$-dependence of the spin
fluctuations in the spin liquid phase of CdCr$_2$O$_4$ measured
above $T_N$. The ring-shaped
intensity around (2,2,0) is essentially identical to the one observed 
in ZnCr$_2$O$_4$, and it is due to the collective
low energy excitations of antiferromagnetic hexagonal spin clusters
in the pyrochlore lattice \cite{lee02}. This suggests that the
cubic phase of CdCr$_2$O$_4$ can be well represented by the Hamiltonian ${\cal H}_1$ as in ZnCr$_2$O$_4$. $J$ is estimated from the Curie-Weiss
temperature $|\Theta_{CW}|$ = 88.97 K (see Fig. 2(a)), to be
$J=-1.02$ meV. Although CdCr$_2$O$_4$ appears to have the same fundamental spin degrees of freedom in the cubic phase as in ZnCr$_2$O$_4$, 
it exhibits strikingly different behaviors in the tetragonal phase below $T_N$ = 7.8 K. It undergoes an {\it
elongation} along the $c$ axis (Fig. \ref{data1}(b) and (d)) and the magnetic long range order has
an {\it incommensurate} characteristic wave vector (see Fig. \ref{data1}(c)).

\begin{figure}
 \includegraphics[width=3.0in]{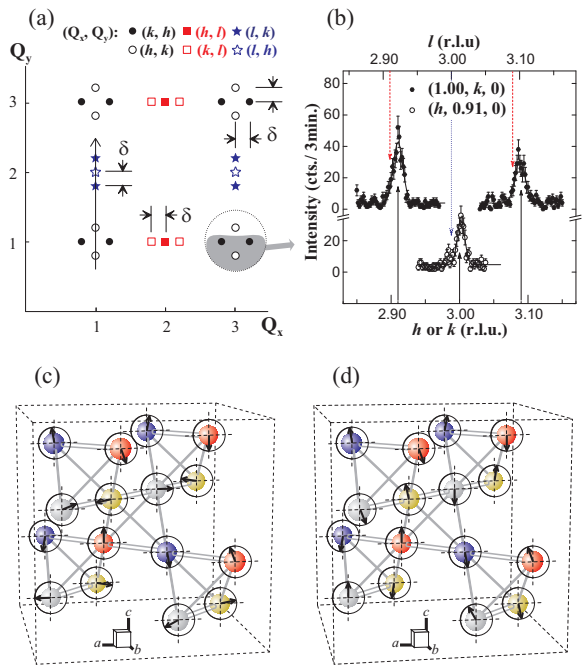}
 \caption{\label{spins} (a) The observed pattern of the magnetic
 reflections on the scattering plane. The quadruplets consist of two doublets at \textbf{Q}$_N\pm$(0,$\delta$,1) and \textbf{Q}$_N\pm$($\delta$,0,1), respectively. The central peaks of the triplets are tails of the doublets that are present above and below the
scattering plane. The colored symbols distinguish the
crystallographic domains. (b) Elastic \textbf{Q}-scans centered at the three peaks under a grey shadow in (a). 
(c),(d) two
possible spin structures deduced from \textbf{Q}$_M$ =
(0,$\delta$,1). Spins are rotating on the $ac$ plane. The double and single lines represent the nearest neighbor bonds in the basal $ab$ plane with $J_{ab}$ and in the out-of-plane with $J_c$, respectively. In (c) the NN spin orientations along $J_c$ are close to perpendicular while in (d) they are close to collinear.
}
\end{figure}

The positions of the magnetic Bragg reflections found in the
scattering plane are shown in Fig. \ref{spins}(a).  The possible characteristic incommensurate wave vectors are
\textbf{Q}$_M$ = (0,$\delta$,1) or (0,1,$\delta$) or
(1,$\delta$,0) with $\delta=0.0894(3)$. ($h$ and $k$ are interchangeable.) However, since $c > a$, the three possible \textbf{Q}$_M$'s would produce 
the magnetic Bragg reflections at slightly different positions in
the scattering plane. We performed elastic scans with a high
Q-resolution over a set of three IC positions to distinguish between the
different scenarios. Fig. 3(b) shows the results. The black solid,
blue dotted and red dashed arrows correspond to the expected peak
positions for \textbf{Q}$_M$ = $(0,\delta,1), (1,\delta,0)$, and $(0,1,\delta)$, respectively. 
All three of the experimental peak positions are consistent with \textbf{Q}$_M$ = (0,$\delta$,1), which indicates that the incommensurability occurs either along the $a$- or the $b$-axis, perpendicular to the elongated $c$-axis. 
Furthermore, we have performed polarized neutron diffraction at TAS1 of JAERI (the details and results of the experiment will be reported elsewhere \cite{kaku05}), and found that the spins are
lying on the plane that is perpendicular to the incommensurability
direction. For convenience, we chose \textbf{Q}$_M$ =
(0,$\delta$,1) with the spins lying in the
$ac$-plane. The resulting breakouts of the magnetic Bragg reflections into three crystallographic and two magnetic domains are shown as different symbols in Fig. 3 (a).

Following group theory arguments \cite{izyu}, the magnetic lattice with the characteristic
wave vector of \textbf{Q}$_M$ consists of four independent
sublattices as represented by spheres with different colors in Fig.
\ref{spins}(c) and (d). Each sublattice connects every
third nearest neighboring Cr$^{3+}$ ions that are separated by the
symmetrically equivalent distance of $(\frac{1}{2}$,$\frac{1}{2}$,0$)$. These are the second
nearest neighbors along the chains represented by lines
in Fig. \ref{spins}(c) and (d). Within the sublattice,
spins are aligned according to $\textbf{S}_j = \textbf{S}_oe^{2\pi
i\textbf{Q}_M \cdot (\textbf{r}_j-\textbf{r}_0)}$ and therefore
the neighbour spins rotate by $2\alpha$, $\pi$, and
$2\alpha+\pi$ in the $ab$-, $ac$- and $bc$-plane, respectively,
where $2 \alpha=\delta\cdot \pi = 16.2^o$. In order to
construct the relative orientation between the different
sublattices, the magnetic interactions in the
tetragonal phase of CdCr$_2$O$_4$ are considered. Analysis of a series of
chromium oxides indicates that $dJ/dr \approx 40$ meV/\AA\
\cite{moti70}. This implies that the tetragonal distortion
($c>a=b$) yields stronger AFM interactions in the basal plane with
$J_{ab}= -1.19$ meV (double lines in Fig. 3 (c) and (d)) and weaker AFM interactions among all other
spin pairs with $J_{c}=-0.95$ meV (single lines). For each chain in the basal plane, the nearest neighboring spins that belong to two different sublattices would favor a phase difference of $\pi \pm \alpha$ to minimize the exchange energy due to the strong AFM $J_{ab}$. The
out-of-plane $J_c$ is frustrating that allows stacking of the $<110>$ chains along the $c$ axis. Fig. \ref{spins}(c)
and (d) show two highly symmetric spin structures that have the same
mean-field exchange energy; in one, the chains are stacked almost orthogonally along the $c$ axis (Fig.
\ref{spins}(c)) and the other, the chains are stacked almost collinearly (Fig. \ref{spins}(d)).

\begin{figure}
 \includegraphics[width=3.6in]{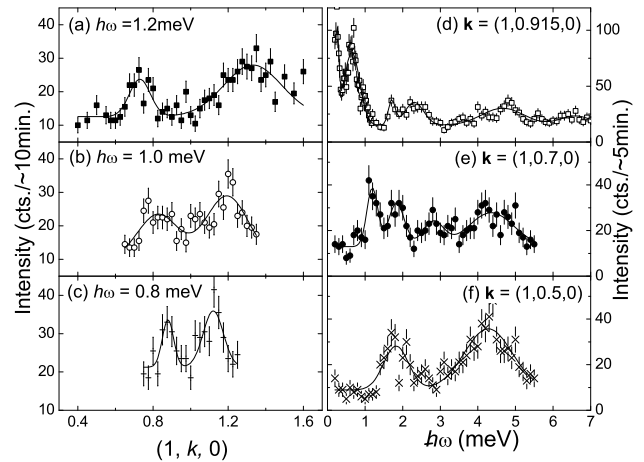}
 \caption{\label{cuts} Constant-$\hbar\omega$ ((a)-(c)) and constant-\textbf{Q}
 ((d)-(e))
 scans of the spin wave excitations in CdCr$_2$O$_4$. The solid lines are guide to eyes.}
\end{figure}

\begin{figure}
 \includegraphics[width=3.6in]{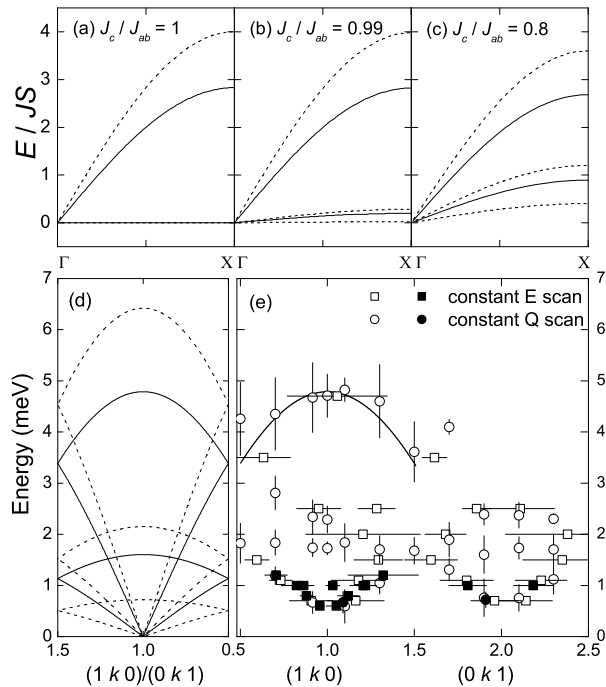}
 \caption{\label{disp} (a), (b), (c), (d) The spin wave calculations with the orthogonal (solid lines) and the collinear (dashed lines) commensurate models with different ratios of $J_{c}/J_{ab}$. In (a), (b), and (c) the energy is in unit of $J_{ab}S$ while in (d) the energy is calculated in meV with $J_{ab}$ = $-1.19$ meV and $J_c=-$0.95 meV. Since the overlap between the (1,$k$,0) and (0,$k$,1) domains is considered, the $k$ = 1 point corresponds to the AFM zone center in the former case while it corresponds to the zone boundary for the latter. 
(e) The dispersion relations determined from constant-\textbf{Q} and contant-$\hbar\omega$ scans. The line is guide to eyes.
The data were taken along the direction in the \textbf{Q} space shown as an arrow in Fig. \ref{spins}(a).
The open and closed symbols are the data obtained with focusing
and flat analyzers, respectively.}
\end{figure}
To understand the elementary excitations of
the IC N\'{e}el state, constant-$\hbar\omega$ and contant-{\bf Q} scans were performed (Fig. \ref{cuts}).
The constant-$Q$ scan
at an IC zone center (Fig. \ref{cuts}(d)) shows that
there is a gap, $\Delta$, of 0.65 meV along with at least two
additional excitation peaks around 2.3 and 4.7 meV. At {\bf Q} values away from the zone center, the lowest energy peak shifts considerably in energy while the two higher energy peaks shift only slightly (Fig. \ref{cuts}
(e)-(f)). Fig. \ref{cuts} (a)-(c) show the dispersiveness at low energies. 
Additional scans were performed to map out the dispersion relation of
the magnetic fluctuations along the (1,$k$,0) direction (the arrow from \textbf{Q}$_y =$ 0.5 to 2.5 
in Fig. \ref{spins}(a)), summarized in Fig. \ref{disp}(e).

What kind of spin Hamiltonian would select the observed IC N\'{e}el state as the ground state of CdCr$_2$O$_4$ in the tetragonal phase? The exchange anisotropy of the nearest neighboring (NN) couplings, $J_c \neq J_{ab}$, alone would favor a commensurate spin structure rather than the IC one, which means that additional perturbations to ${\cal H}_1$ are present. However, $\Delta\simeq 0.6$ meV 
is small compared to the entire energy band width of the dispersion, $\sim$ 5 meV, that suggests that the additional perturbations must be small. Let us first discuss the effect of the exchange anisotropy of the NN couplings, $J_c \neq J_{ab}$, on the dispersion of the two model spin structures shown in Fig. 3 (c) and (d). 
In the ideal isotropic case, $J_c = J_{ab}$, there
is a four-fold (six-fold) degenerate flat zero energy mode and another four-fold (two-fold) degenerate
dispersive mode that extends to higher energies in the
orthogonal (collinear) model (Fig. \ref{disp}(a)). As the anisotropy is introduced, the zero-energy mode becomes dispersive. In the collinear structure, the six-fold zero-energy
mode splits into two-fold and four-fold weakly dispersive
modes, whereas it remains
four-fold degenerate for the orthogonal structure \cite{110}. The dispersions are not affected by the exchange anisotropy at high energies as much as they do at low
energies. Indeed, other anisotropy terms such as the single ion anisotropy
and the Dzyaloshinsky-Moriya (DM) interactions also modify the dispersion at low energies considerably
but not at high energies. Fig. \ref{disp}(d) shows the
dispersion relations obtained using the experimentally determined $J_{ab} = -1.19$ meV and $J_c = -0.95$ meV.
This shows that the orthogonal spin structure fits the dispersion at high
energies better than the collinear one. This means that the N\'{e}el state of CdCr$_2$O$_4$ is likely to be the incommensurate spin structure
with the orthogonal stacking shown in Fig. \ref{spins}(c). The actual incommensurability seems to be caused by other
perturbative terms, such as further
nearest interactions and/or DM interactions. We also considered the exchange interactions between third nearest neighbor interactions, and extensively examined the phase space as a function of $J$. We found a region where an incommensurate spin structure can be selected as a ground state but the incommensurability was along the $<110>$ direction, as recently observed in LiCuVO$_4$ \cite{lcvo}, rather than the observed $<010>$ direction.
This suggests that the tetragonal distortion involves distortions of the oxygen octahedra 
that lowers the crystal symmetry. 

In summary, we identified a spin-lattice coupling mechanism
that lifts the magnetic frustration in CdCr$_2$O$_4$ and that is distinctly different from the one observed in ZnCr$_2$O$_4$. The tetragonal distortion involves an elongation along the $c$-axis and the N\'{e}el state has a helical spin structure with the single characteristic wave vector of \textbf{Q}$_M$ = (0,$\delta$,1). 
Our identification of the spin structure and dynamics of the low temperature phase of CdCr$_2$O$_4$ should provide a unique test to theoretical attempts to explain the spin-Peierls-like phase transitions in the Heisenberg pyrochlore 
antiferromagnets.

\begin{acknowledgments}
We thank C. L. Henley, Y. Motome, A. Zheludev, and D. Louca for valuable
discussions. This work was partially supported by the NSF through
DMR-9986442 and by the U.S. DOC through NIST-70NANB5H1152.
\end{acknowledgments}



\begin{thebibliography}{}
\bibitem{110} Note that the two spin structures with $\delta=0$ produce exactly the same spectrum along the (110) chain directions.
\end{thebibliography}
\end{document}